\documentclass[pra,twocolumn,showpacs,amsfonts,superscriptaddress,amsmath,floatfix]{revtex4}
\usepackage{graphicx}
\usepackage{bm}
\usepackage{color}

\begin{document}

\title {Quantum superpositions of flow states on a ring}
\author{Rina Kanamoto}
\email{kanamoto.rina@ocha.ac.jp}
\address{Division of Advanced Sciences, Ochanomizu University, Bunkyo-ku, Tokyo, 112-8610, Japan}

\author{Ewan M. Wright}
\email{ewan.wright@optics.arizona.edu}
\address{College of Optical Sciences, University of Arizona,
Tucson, AZ 85721, USA}

\date{\today}


\begin{abstract}
We propose a scheme to generate quantum superpositions of macroscopically distinct flow states of ultracold atoms on a ring using Raman coupling employing a quantized laser field that is a cat-like superposition of optical vortices with opposite winding numbers. For atoms initially in their ground state and entangled optical vortices of coherent states, this scheme can produce a superposition of rotating and non-rotating states.  We find fidelities of the quantum superpositions around $0.9$ even before optimization.
\end{abstract}
\maketitle

\section{Introduction}

It has been stressed by Leggett \cite{Leg02} that quantum superpositions of macroscopically distinguishable states are invaluable for testing the validity of quantum theory as well as advancing emerging quantum technologies such as precision measurement and quantum computing. A prime example is the observation of quantum superpositions of persistent currents in superconducting quantum interference devices that are candidates as qubits in future quantum computers~\cite{NakPasTsa99,FriPatChe00,WalHaaWil00}. Spurred on by these developments there are several proposals to create flow states in atomic Bose-Einstein condensates (BECs), either in a trap or on a ring.
Flow states are those states in which all atoms are in the same angular-momentum eigenstate with winding number $q$ so that
the total angular momentum is $\hbar Nq$, and a BEC in a flow state exhibits the peculiarities of superfluidity.
Proposals include stirring using rotating light-shift potentials \cite{MarZha98a}, ``phase engineering'' involving a Gaussian laser beam whose center is rotated and that couples the external motion to the internal state via Rabi oscillations \cite{MatAndHal99,WilHol99}, and most significantly for this work, vortex coupling in which two-photon stimulated Raman transitions are driven using Laguerre-Gaussian (LG) fields to transfer orbital angular momentum (OAM) from the LG beam photons to the trapped atoms \cite{MarZhaWri97,KapDow05}. Experiments from NIST have already demonstrated quantized rotation of trapped atoms using LG fields \cite{AndRyuCla06}, and persistent flow of a BEC in a toroidal trap \cite{RyuAndCla07}.

The above proposals do not produce quantum superpositions of flow states but rather {\it each atom} is generally excited into a superposition of angular-momentum states.  In contrast theoretical works have appeared involving creation of macroscopic superpositions of BECs \cite{CirLewMol98,SorDuaCir01,DunHal06,DunBurRot06,NunReyBur08}.  Of particular interest here is the recent work on macroscopic superposition states in ring superlattices in which an array of BECs trapped in optical potentials are coupled via tunneling and formed into a ring, thereby creating a discrete analogue of a ring BEC \cite{DunHal06,DunBurRot06,NunReyBur08}.  These authors have proposed a scheme for creating macroscopic superpositions of persistent flows on the ring superlattice.


\begin{figure}[t]
\includegraphics[scale=0.5]{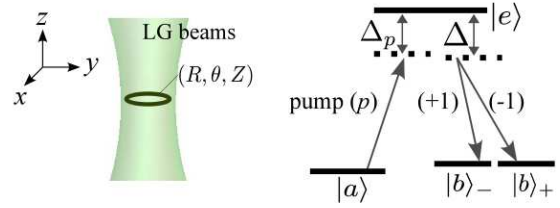}
\caption{
Ultracold atoms trapped on a ring are driven by two co-propagating LG beams.
The atoms are initially in the non-rotating hyperfine ground state $|a\rangle$ are driven
via two-photon stimulated Raman transitions to different hyperfine ground states $|b\rangle_{\pm}$
with different winding numbers via electronic excited state $|e\rangle$.
The winding numbers of the LG fields involved are shown in parenthesis.
}
\label{fig1}
\end{figure}

The goal of this article is to propose a new scheme to generate quantum superpositions of macroscopically distinct flow states for ultracold bosonic atoms trapped circumferentially on a ring, as opposed to a ring lattice, using two-photon stimulated Raman transitions driven by a pump field that is a classical optical vortex, and a second quantized laser field that is a cat state of optical vortices of opposite winding number $\ell=\pm 1$. Assuming that all the atoms are initially in a non-rotating hyperfine ground state atomic level, the key to our scheme is that the combination of the classical pump field with each of the two individual components of the cat state can drive the atoms via the two-photon transitions into different hyperfine ground states representing different flow states, so that the field cat state can generate a quantum superposition of flow states. Wright has previously described how cat states of optical vortices may be generated, and introduced an idea of generating macroscopic superpositions of matter wave as a potential application of an optical cat state~\cite{Wri09}. In this article we advance this idea in a concrete geometry and explicitly show that the approach is feasible. That the cat state nature of the driving quantized optical field can be transferred to the atoms is not surprising in view of previous studies of quantum atom optics involving BECs in quantized light fields~\cite{CirLewMol98,MooZobMey99}, and of conditional quantum dynamics, in which the quantum dynamics of a subsystem, here the atoms, depends on the state of another subsystem, here the light field, wherein the initially uncorrelated two subsystems get entangled~\cite{Haroche94,Jozsa95}. In the remainder of this article we first discuss cat states of optical vortices that are required for our scheme and then turn to the analysis of how the quantum superpositions of flow states are generated.


\section{Optical cat states}

For our scheme the atoms are assumed to be trapped
circumferentially on a ring in the $x$-$y$ plane and to be illuminated by
a pair of laser fields co-propagating along the $z$-axis whose centers coincide with
the ring center, see Fig.~\ref{fig1}.  The pump field is a LG beam of winding number $p$ with
$|p| \gg 1$ that we will treat classically, and the second field is a cat state of LG fields
of winding numbers $\ell=\pm 1$ that we treat quantum mechanically.  In particular, taking
the cat state and the pump laser to have frequencies $\omega$, $\omega_{p}$,
polarizations ${\bf e}_{LP}$, ${\bf e}_{P}$, mode profiles $u_1(r)$, $u_{p}(r)$,
mode volumes $V_{LP}$, $V_{P}$, and propagation numbers $\beta_1$, $\beta_{p}$,
respectively, we write the positive frequency component for
the entire light field operator in the Heisenberg picture as
$\hat{\bf E}^{(+)}({\bf r},t)=i\sqrt{\frac{\hbar \omega_{p}}{2\epsilon_{0 V_{P}}}}{\bf e}_{P}
u_p(r) e^{i\beta_{p} z}\hat{c}_{p}e^{ip\theta}+i\sqrt{{\hbar\omega\over2\epsilon_0V_{LP}}}{\bf e}_{LP}
u_1(r)e^{i\beta_1z}\left [\hat c_+(t)e^{i\theta} +\hat c_-(t)e^{-i\theta}\right ]$.
Here $\hat c_\pm$ and $\hat c_{p}$ are the annihilation operators for the LG modes with winding numbers
$\ell=\pm 1$ and $p$. Furthermore, we consider cat states of optical vortices of the form
$|\Phi_{\rm field}(\phi)\rangle \propto  [|\phi\rangle_+|0\rangle_-+|0\rangle_+|\phi\rangle_-]$,
where $|\phi\rangle_{\pm}$ represents the occupation of the OAM state $\ell = \pm 1$ in
a quantum state $\phi$, and $|0\rangle_{\pm}$ the vacuum.
A natural choice is a cat states of coherent states with opposite winding numbers $|\phi\rangle_\pm=|\alpha\rangle_\pm$,
\begin{eqnarray}\label{coherent_cat}
|\Phi_{\rm field}(\alpha)\rangle
=[|\alpha\rangle_+|0\rangle_-+|0\rangle_+|\alpha\rangle_-]/\sqrt{2}.
\end{eqnarray}
In writing this field quantum state we have assumed that $|\alpha|^2 \gg 1$ so that
we may treat $|\alpha\rangle$ and the vacuum state $|0\rangle$ as orthogonal for all practical purposes.
We thus consistently neglect all terms of order $e^{-|\alpha|^2}$ throughout.

Some words are in order regarding the generation of cat states of optical vortices.
A great deal of research has been directed towards creating quantum superpositions of optical vortices
for use in quantum information processing, but these typically involve small numbers of photons~\cite{ArnBar00,MaiVazWei01,TerTorTor02}.
Wright has previously described how cat states of optical vortices may be generated using
self-phase modulation based on electromagnetically-induced transparency enhanced Kerr nonlinearities
for atoms loaded into hollow core photonic-crystal fibers \cite{Wri09}, and Glancy and Macedo de Vasconcelos~\cite{GlaVas08}
have reviewed a number of methods to produce cat states of coherent states, including the Kerr effect, degenerate optical parametric oscillators,
back-action evasion measurement, and photon subtraction. Furthermore, recent work by De Martini and co-workers \cite{MarSciVit08} shows promise for creating an entangled field state with high resilience against the detrimental effects of decoherence \cite{DeMSciSpa09} in which a macrostate with more than $10^4$ photons is entangled with a microstate containing one photon using a quantum-injected quantum parametric amplifier. Each of these methods could be used to create cat states of optical vortices if the two coherent state components of the cat state can be produced with suitable entangled spatial beam properties \cite{JanWagMor09}.


\section{Raman coupler}

Figure~\ref{fig1} shows the configuration of the hyperfine atomic states
denoted as $|a\rangle$, $|b\rangle_{\pm}$, and $|e\rangle$, coupled via the LG fields. The atoms are initially prepared in the non-rotating state $|a\rangle$, the pump laser
couples the initial state $|a\rangle$ and an excited state $|e\rangle$ of angular momentum $p$, while
the quantized light field in the optical cat state couples $|b\rangle_{\pm}$ of the angular momentum $p\pm 1$ and $|e\rangle$, transitions between $|a\rangle$ and $|b\rangle_{\pm}$ being dipole forbidden.
The single-photon detunings $\Delta_p=[\omega_p-(\omega_e-\omega_a)],\Delta=[\omega-(\omega_e-\omega_b)]$ are chosen
large enough that the excited state $|e\rangle$ is negligibly populated by single-photon absorption, allowing the excited state to be adiabatically eliminated. Here we denote the energy level of each hyperfine state as $\hbar\omega_j=\int drdz \psi_0^*\left[-\frac{\hbar^2}{2M}\nabla_{r,z}^2 + V_j(r,z)+\epsilon_j\right] \psi_0$ with $V_j$ and $\epsilon_{j}$ $(j=a,b,e)$ being the single-particle trapping potentials and the internal energies, respectively, and $M$ is the atomic mass. We assume that the atoms are tightly trapped circumferentially on a ring by the single-particle potential, and that they are described by the same cylindrically symmetric atomic center-of-mass ground state wave function $\psi_{0}(r,z)$. Then treating the pump laser as a classical $c$-number field $\hat{c}_p \to \alpha_p$, the Hamiltonian for our system in the dipole and rotating-wave approximations reads
\begin{widetext}
\begin{eqnarray}\label{H}
\hat{H}
&=&
\left[ \hbar\delta +\frac{\hbar^2(p+1)^2}{2MR^2} \right]\hat{b}_+^{\dagger}\hat{b}_+
+\left[ \hbar\delta +\frac{\hbar^2(p-1)^2}{2MR^2} \right]\hat{b}_-^{\dagger}\hat{b}_-
+\hat{H}_{\rm col}
+\left[\hbar g\alpha_p^*\hat{a}^{\dagger}(\hat{b}_-\hat{c}_++\hat{b}_+\hat{c}_-)+h.c.\right]\nonumber\\
&&+\left[\hbar g_{bc}\hat{b}_-^{\dagger}\hat{b}_+\hat{c}_+^{\dagger}\hat{c}_-+h.c.\right]
+\hbar g_{bc}(\hat{c}_+^{\dagger}\hat{c}_++\hat{c}_-^{\dagger}\hat{c}_-)
(\hat{b}_+^{\dagger}\hat{b}_++\hat{b}_-^{\dagger}\hat{b}_-)+\hbar g_{ap}|\alpha_p|^2 \hat{a}^{\dagger}\hat{a}
\end{eqnarray}
\end{widetext}
where $\delta=(\Delta-\Delta_p)=(\omega-\omega_p)+(\omega_b-\omega_a)$ denotes the two-photon detuning.
The atom-field coupling constants are given by
$\hbar g= \frac{d_{be}d_{ea}}{\epsilon (\Delta+\Delta_p)}\sqrt{\frac{\omega \omega_p}{V_PV_{LP}}}
\int drdz |\psi_0|^2 \psi_p^*\psi$,
$\hbar g_{bc}=\frac{|d_{be}|^2\omega}{\epsilon (\Delta+\Delta_p) V_{LP}}
\int drdz |\psi_0|^2|\psi|^2$, and
$\hbar g_{ap}=\frac{|d_{ae}|^2\omega_p}{\epsilon (\Delta+\Delta_p) V_P}
\int drdz |\psi_0|^2|\psi_p|^2$
with $\psi(r,z)=u_1(r)e^{i\beta_1 z}$, $\psi_p (r,z)=u_p(r) e^{i\beta_p z}$,
and $d_{jj'}$ being the dipole matrix elements.
The terms $\hbar (p \pm 1)^2/(2MR^2)$ correspond to the kinetic energy (external momentum)
of the internal state $|b\rangle$ derived from the angular momentum conservation $0 \to p \to p \pm 1$ via the two-photon transition.
The corresponding atomic wavefunctions with the winding number $l$ are written as $\psi_{l}(r,z,\theta)=\psi_0(r,z)e^{il\theta}/\sqrt{2\pi}$,
where $l=0$ for $|a\rangle$, $l=p \pm 1$ for $|b\rangle_{\pm}$, and $\psi_0(r,z)$ being the ground-state mode profile of the ring trap.
The term $\hat{H}_{\rm col}$ denotes atom-atom interactions
\begin{eqnarray}
\hat{H}_{\rm col}\!\!\!&=&\!\!\!
\hbar {G}_{aa}\hat{a}^{\dagger}\hat{a}^{\dagger}\hat{a}\hat{a}
+\sum_{s=\pm }\left[\hbar {G}_{ab}\hat{a}^{\dagger}\hat{a}\hat{b}_s^{\dagger}\hat{b}_s\right.\nonumber\\
&&\left.\!\!\!\!\!\!+\hbar G_{bb}(\hat{b}_s^{\dagger}\hat{b}_s^{\dagger}\hat{b}_s\hat{b}_s
+\hat{b}_s^{\dagger}\hat{b}_s^{\dagger}\hat{b}_{-s}\hat{b}_{-s}
+2\hat{b}_s^{\dagger}\hat{b}_s\hat{b}_{-s}^{\dagger}\hat{b}_{-s})
\right]
\end{eqnarray}
where $\hbar {G}_{jj}=\hbar^2l_{jj} M^{-1}\int drdz |\psi_0(r,z)|^4$ ($j=a,b$)
and $\hbar {G}_{ab}=2\hbar^2 l_{ab} M^{-1}\int drdz |\psi_0(r,z)|^4$
stand for the strength of atomic interactions with the $s$-wave scattering lengths $l_{jj'}$.

We consider an initial quantum state with the light field in the cat state Eq. (\ref{coherent_cat}),
and the initial atomic state $|N,0,0\rangle$ in terms of Fock basis $|n_a,n_+,n_-\rangle$ with
respect to the atomic states $|a\rangle,|b\rangle_+,|b\rangle_-$.
We assume that atoms and field are initially uncorrelated for $t<0$, and the initial state is thus written as
$|\Psi(t=0)\rangle=|N,0,0\rangle|\Phi_{\rm field}(\alpha)\rangle$.
When the atoms are irradiated by the pump and cat state fields they will drive two-photon transitions between the levels $a$
and $b_{\pm}$, and the atoms will in turn re-radiate and produce new radiation in addition to the incident fields, and in this way
the fields will generally be time-dependent.  However, our main focus here is the quantum dynamics of the atoms as opposed to
the radiation produced by the atoms, and we make the approximation that the fields acting specifically on the atoms may be approximated
by the unmodified incident fields.  This is very reasonable given that any time-dependent radiated fields will be directed away from
the ring and our geometry does no include an optical cavity that could redirect the radiated fields back onto the ring.
The combined quantum state therefore evolves for $t>0$ as
\begin{eqnarray}\label{Psi}
|\Psi(t)\rangle={1\over\sqrt{2}}\left[|\Phi_{\rm atom}^{(1)}(t)\rangle|\alpha\rangle_+|0\rangle_-
+ |\Phi_{\rm atom}^{(2)}(t)\rangle|0\rangle_+|\alpha\rangle_-\right],
\end{eqnarray}
where $|\Phi_{\rm atom}^{(1,2)}(t)\rangle$ denote atomic states associated with
the components of the field cat state $|\alpha\rangle_+|0\rangle_-$ and $|0\rangle_+|\alpha\rangle_-$ which
represent quasi-classical fields with well defined winding numbers $\ell=\pm 1$, respectively.
Thus, when the atomic states associated with the two components $|\Phi_{\rm atom}^{(1)}\rangle$ and $|\Phi_{\rm atom}^{(2)}\rangle$ are macroscopically distinguishable, Eq.~(\ref{Psi}) represents a Schr\"odinger cat state.


\begin{figure}[b]
\includegraphics[scale=0.55]{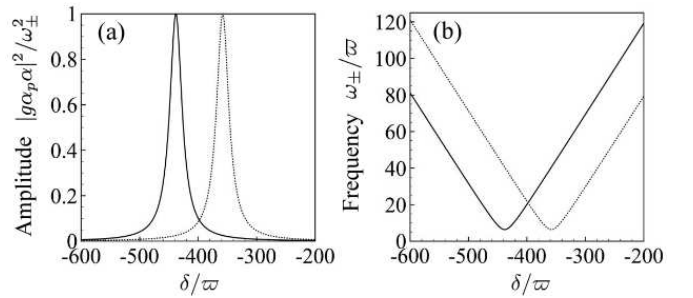}
\caption{
Amplitude and frequency of Rabi oscillation between $|a\rangle$--$|b_-\rangle$ (solid curves), and
$|a\rangle$--$|b_+\rangle$ (dotted curves) when the light field with the OAM $\ell=1$ or
$\ell=-1$ is respectively turned on. The atom-field couplings are $g=g_{ap}=g_{bc}=8\times 10^{-2} \varpi$,
the OAM of the classical pump laser $p=-20$. We took the coherent amplitudes as real numbers
without loss of generality, $\alpha_p=10$ and $\alpha=8$.}
\label{fig2}
\end{figure}


\section{Generation of matter-wave superposition}

We start with examination of the quantum dynamics with just one component,
say $\ell=+1$, of the optical cat state present, and we neglect atomic collisions for simplicity.
Replacing $\hat{c}_+\to \alpha$, $\hat{c}_-=0$
in the Hamiltonian~(\ref{H}), we obtain the Heisenberg equations of motion
$id\hat{a}/dt=g_{ap}|\alpha_p|^2\hat{a}+g\alpha_p\alpha\hat{b}_-$,
$id\hat{b}_+/dt=\Omega_+\hat{b}_+$, and
$id\hat{b}_-/dt=\Omega_-\hat{b}_-+g^*\alpha_p^*\alpha^*\hat{a}$,
where $\Omega_{\pm}=[\delta+\varpi(p\pm 1)^2+g_{bc}|\alpha|^2-g_{ap}\alpha_p^2]/2$, and $\varpi \equiv \hbar/(2MR^2)$.
The above Heisenberg equations are exactly soluble, and for the initial atomic state $|N,0,0\rangle$,
the population in each component evolves in time as
\begin{eqnarray}
&&N_a(t)=N\left[\cos^2 (\omega_- t) + \left(\frac{\Omega_-}{\omega_-}\right)^2\sin^2(\omega_- t)\right]\\
&&N_+(t)=0,\quad N_-(t)=N\left (\frac{|g\alpha_p\alpha|^2}{\omega_-^2}\right )\sin^2 (\omega_- t) ,
\end{eqnarray}
with the Rabi frequency $\omega_{-}=\sqrt{\Omega_{-}^2+|g\alpha_p\alpha|^2}$.
At the two-photon detuning $\delta^{\rm (1)}=-\varpi (p-1)^2-g_{bc}|\alpha|^2+g_{ap}\alpha_p^2$, complete and periodic oscillations occur between the non-rotating state $|a\rangle$ and the atomic state $|b\rangle_-$ with winding number $(p-1)$.
Similarly, if we consider solely the $\ell=-1$ component of the field cat state
one finds complete periodic oscillations of frequency $\omega_+$ between the non-rotating state $|a\rangle$ and the atomic state $|b\rangle_+$ with winding number $(p+1)$ for a detuning $\delta^{\rm (2)}=-\varpi (p+1)^2-g_{bc}|\alpha|^2+g_{ap}\alpha_p^2$. Figure~\ref{fig2} shows (a) the maximum transfer amplitudes $(N_\pm/N)$ of the oscillations between the two atomic states involved, and (b) the oscillation frequencies $\omega_\pm$ normalized by $\varpi$ for the two components of the field cat state, both as functions of detuning $\delta$.

The crux of our proposal is now as follows: Since the two transfer amplitude curves in Fig.~\ref{fig2}(a) are well resolved this means that by choosing the detuning $\delta$ at one of the peaks we can have one of the atomic components $|\Phi_{\rm atom}^{(1,2)}\rangle$ be off-resonant and remain largely in the non-rotating state, whereas the other atomic state component $|\Phi_{\rm atom}^{(2,1)}\rangle$ is transferred to the flow state with winding number $(p\pm 1)$ depending on the peak chosen. Using this we can generate a quantum superposition as in Eq. (\ref{Psi}), and this is a key message of this paper.

\begin{figure}
\includegraphics[scale=0.6]{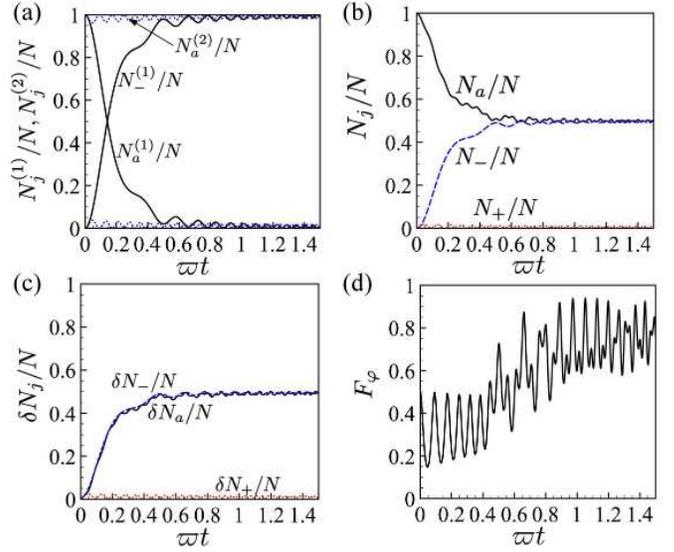}
\caption{
Time evolution of (a) the populations in each atomic branch,
(b) the total populations, (c) number fluctuations, and (d) the fidelity.}
\label{fig3}
\end{figure}

Next we provide a numerical example to demonstrate that our scheme for the generation of quantum superpositions of flow states can survive the inclusion of atomic collisions. Many-body collisions complicate the generation process in that they nonlinearly modify the detuning required for coupling between the atomic states, and the transfer between the atomic states is generally no longer complete.  As previously discussed \cite{MarZhaWri97,KapDow05} this can be offset by sweeping the detuning in time, and we do this here by setting $\delta(t)/\varpi =-420-95 \varpi t$ so that the detuning at $t=0$ is to the right of the left peak in Fig.~\ref{fig2}(a). We set $N=64$, $G_{aa}=G_{bb}=8\times 10^{-3} \varpi$, and $G_{ab}=2\times 10^{-3} \varpi$, all other parameters being identical to those in Fig.~\ref{fig2}. The atomic state vector is expanded in terms of the Fock basis
$|\Phi_{\rm atom}^{(1,2)}(t)\rangle = \sum_{\{n\}} A_{\{n\}}^{(1,2)}(t)|n_a,n_+,n_-\rangle$ where
$\{n\}$ signifies summing over all combinations of the three zero or positive integers $\{n_a,n_+,n_-\}$ satisfying the condition
$N=n_a+n_++n_-$, and the equations of motion for the amplitudes $A_{\{n\}}^{(1,2)}(t)$ according to the Hamiltonian (\ref{H}) are solved numerically. Figure~\ref{fig3} shows (a) the population in each atomic component $N_j^{(1,2)}=\langle \Phi^{(1,2)}_{\rm atom}|\hat{N}_j |\Phi^{(1,2)}_{\rm atom} \rangle$ with $j=a,\pm$, (b) the population in each atomic state
$N_j=\langle \Psi|\hat{N}_j |\Psi\rangle$, and (c) the corresponding number fluctuations $\delta N_j =\sqrt{\langle \Psi | \hat{N}_j^2 |\Psi\rangle -\langle \Psi | \hat{N}_j | \Psi\rangle^2}$, all as a function of time. Figure~\ref{fig3}(a) demonstrates that since the detuning is such that the atomic state $|\Psi_{\rm atom}^{(2)}\rangle$ remains off-resonance, the population is transferred only in the branch $|\Phi_{\rm atom}^{(1)}\rangle$ from $|a\rangle\rightarrow |b\rangle_-$. Furthermore, for $\varpi t>0.6$ Fig.~\ref{fig3}(b) demonstrates that the non-rotating state $|a\rangle$ and the rotating state $|b\rangle_-$ with winding number $(p-1)$ occur with equal probability $0.5$, indicative of the desired quantum superposition but indistinguishable at this point from a number state where half the population is transferred to each component $|\Phi_{\rm atom}\rangle \simeq |N/2,0,N/2\rangle$.
To distinguish these states we realize that we have as a target state the quantum superposition $|\Psi_{\varphi}\rangle=[|0,0,N\rangle|\alpha\rangle_+|0\rangle_- +e^{i\varphi}|N,0,0\rangle |0\rangle_+ |\alpha\rangle_-]/\sqrt{2}$, with $\varphi$ a phase resulting from the light-matter and atom interactions. For the quantum superposition the number fluctuations are $\delta N_a =\delta N_- = N/2$, $\delta N_+ =0$, whereas for the number state $\delta N_a = \delta N_- = N/4$, $\delta N_+=0$, and Figure~\ref{fig3}(c) supports the case that our scheme generates a quantum superposition of flow states.
Finally we introduce the fidelity $F_{\varphi}=|\langle \Psi_{\varphi}|\Psi(t)\rangle|$ of the quantum superposition, and for each time we vary the phase $\varphi$ to attain the maximum value.  The fidelity is plotted versus time in Fig.~\ref{fig3}(d) and we find a maximum value $F_{\varphi}\simeq 0.9$, impressive given that we have not applied any particular optimization. The reduction of the fidelity from the ideal value of unity is associated with the detrimental effects of collisions leading to a broad distribution of the atom statistics $|A_{\{n\}}^{(1)}|^2$ that are peaked around $A_{0,0,N}^{(1)}$. It thus follows that increasing collisions tends to reduce the peak fidelity, so reducing collisions is one strategy, but we anticipate that the fidelity can be greatly improved even for larger collisions by applying quantum control methods to our scheme, and we plan to pursue this in future work.


\section{Conclusion}

We have described a new scheme to generate quantum superpositions of flow states of ultracold atoms on a ring using vortex coupling employing a coherent state optical vortex cat state, and numerically demonstrated that the scheme works even in the presence of collisions. An outstanding issue is the extreme fragility of coherent state cat states to decoherence which is an issue for even generating the field cat state.  To address this we also performed simulations employing squeezed state optical vortex cat states, which have increased resilience to decoherence~\cite{SerDeSIll04}, and found very similar results to Fig.~\ref{fig3}, with almost identical maximum fidelities. There is also the issue of how our results are modified in the limit of cat field states with low photon number which are more resilient to decoherence.  In that case our analysis becomes very similar to that of Lo Gullo {\it et al.,} who considered the entanglement of spatially separated BECs using OAM-entangled two-photon states, and their analysis illustrates that entanglement still occurs for small photon numbers beyond our approximations~\cite{LoGullo10}. We plan to explore these and other issues further in future publications.

RK was supported by a Grant-in-Aid for Scientific Research under Grant No. 21710098 and the Sumitomo Foundation. EMW is supported in part by the Joint Services Optical Program (JSOP).


%
\end{document}